\documentclass[11pt]{article}

\usepackage{bm, commath}
\usepackage{natbib}
\usepackage{caption}
\usepackage{graphicx}
\usepackage{subcaption}
\usepackage{amsmath, amsfonts, amsthm}
\usepackage{float}
\usepackage{booktabs,siunitx}
\usepackage{url}
\usepackage{multirow}
\usepackage{amssymb}
\usepackage{bm}
\usepackage{mathrsfs}
\usepackage{xr}
\usepackage{authblk}

\usepackage{listings}
\allowdisplaybreaks
\usepackage{bbm}
\usepackage{textcomp}
\usepackage[margin=1in]{geometry}
\usepackage{authblk}
\usepackage[ruled]{algorithm2e}
\SetKwInput{KwParam}{Parameter}
\SetAlgoCaptionLayout{centerline}

\usepackage{sectsty}
\usepackage[onehalfspacing]{setspace}
\usepackage[usenames,dvipsnames]{color}
\usepackage[utf8]{inputenc}
\usepackage{float}
\usepackage{tikz}
\usetikzlibrary{trees}
\usetikzlibrary{shapes,decorations,arrows,calc,arrows.meta,fit,positioning}
\usetikzlibrary{shapes.multipart}
\tikzset{
    -Latex,auto,node distance =1 cm and 1 cm,semithick,
    state/.style ={ellipse, draw, minimum width = 0.7 cm},
    state1/.style ={ draw, minimum width = 0.7 cm},
    point/.style = {circle, draw, inner sep=0.04cm,fill,node contents={}},
    bidirected/.style={Latex-Latex,dashed},
    el/.style = {inner sep=2pt, align=left, sloped}
}

\def\var{\mathrm{var}}
\def\cov{\mathrm{cov}}
\def\X{\bm X} 
\def\A{\bm A} 
\def\a{\bm a} 
\def\bbeta{\bm \beta}

\def\btheta{\bm\theta}

\usepackage{mathtools}

\usepackage{xcolor}

\makeatletter
\renewcommand{\algocf@captiontext}[2]{#1\algocf@typo. \AlCapFnt{}#2} 
\def\@algocf@capt@plain{top}
\renewcommand{\algocf@makecaption}[2]{%
  \addtolength{\hsize}{\algomargin}%
  \sbox\@tempboxa{\algocf@captiontext{#1}{#2}}%
  \ifdim\wd\@tempboxa >\hsize
  \hskip .5\algomargin%
  \parbox[t]{\hsize}{\algocf@captiontext{#1}{#2}}
  \else%
  \global\@minipagefalse%
  \hbox to\hsize{\box\@tempboxa}
  \fi%
  \addtolength{\hsize}{-\algomargin}%
}
\makeatother

\begin{document}

\sectionfont{\bfseries\large\sffamily}%

\subsectionfont{\bfseries\sffamily\normalsize}%

\title{Robust Variance Estimation for Covariate-Adjusted Unconditional Treatment Effect in Randomized Clinical Trials with Binary Outcomes}

\author{Ting Ye}
\author{Marlena Bannick}
\affil{Department of Biostatistics, University of Washington,  Seattle, Washington 98195, U.S.A.}

\author{Yanyao Yi}
\affil{Global Statistical Sciences, Eli Lilly and Company, 
	Indianapolis, Indiana  46285, U.S.A.}

\author{Jun Shao}
\affil{School of Statistics, East China Normal University, Shanghai 200241, China\\ Department of Statistics, University of Wisconsin, Madison, Wisconsin 53706, U.S.A.}

\maketitle

\begin{abstract}
	To improve precision of estimation and power of testing hypothesis for an unconditional treatment effect in randomized clinical trials with binary outcomes, researchers and regulatory agencies recommend using g-computation as a reliable method of covariate adjustment. However, 
	 the practical application of g-computation is  hindered by the lack of an explicit robust variance formula that can be used for different unconditional treatment effects of interest. To fill this gap, we provide  explicit and robust variance estimators for g-computation estimators and demonstrate through simulations that the  variance estimators can be reliably applied in practice.

\end{abstract}


\newpage 

\section{Introduction}

In randomized clinical trials, adjusting for baseline covariates has been advocated as a way to improve precision of estimating and power of testing treatment effects {\citep{Yang:2001aa,Tsiatis:2008aa, Freedman:2008ab, Lin:2013aa, ye2022bio, ye2021better}}.  	We focus on binary outcomes in this article.
When a logistic model is used as  a working model for baseline covariate adjustment, 
the g-computation \citep{Freedman:2008ab, Moore:2009tu} provides asymptotically normal estimators of unconditional
treatment effects such as the risk difference, relative risk, and odds ratio, regardless of whether the logistic model is correct or not. 
In May 2021, the US Food and Drug Administration released a draft guidance \citep{fda:2019aa}  for the use of covariates in the analysis of randomized clinical trials, and recommended the g-computation as a ``statistically reliable method of covariate adjustment for an unconditional treatment effect with binary outcomes.'' 

However, to the best of our knowledge, no explicit robust variance estimation formula for  g-computation  is currently available that can be used for inference on different unconditional treatment effects of interest.
 Moreover, some existing variance estimation formulas in the literature, such as the  formula in \cite{ge2011covariate} for risk difference and two treatment arms, are model-based and do not fit the model-robust inference paradigm.  Additionally, the formula in \cite{ge2011covariate} does not take into account a source of variability due to covariates and nonlinearity of logistic model, which can lead to confidence intervals with insufficient coverage probabilities.


 The purpose of this article is to fill this gap by providing explicit and robust variance estimators for g-computation estimators.
 Our simulations demonstrate that the provided  variance estimators can be reliably applied in practice.

\section{Robust Variance Estimation}

Consider a $k$-arm trial with $n$ subjects. For each subject $i$, let $\A_i$ be the $k$-dimensional treatment indicator vector that equals $\a_t$ if patient $i$ receives treatment $t$ for $t=1,\dots, k$, where $\a_t$ denotes the $k$-dimensional vector whose $t$th component is 1 and other components are 0, $Y_i^{(t)}$ be the binary potential outcome under treatment $t$,  and $\X_i$ be the baseline covariate vector for adjustment. The observed outcome is  $Y_i = Y_i^{(t)}$ if and only if $\A_i =\a_t$. We consider simple randomization where $\A_i$ is completely random with known  $\pi_t = P(\A_i = \a_t) $,  $\pi_t >0$ and $\sum_{t=1}^k \pi_t = 1$. We assume that $(Y_i^{(1)},\dots,  Y_i^{(k)}, \A_i, \X_i ),  i=1,\dots, n$, are  independent and identically distributed with
finite second order moments. To simplify the notation, we drop the subscript $i$ when referring to a generic subject from the population. 
Write the unconditional response means as $\theta_t= E(Y^{(t)})$ and $\btheta = (\theta_1, \dots, \theta_k)^T$, where the superscript $T$ denotes transpose of a vector throughout.  
 The target parameter is a given contrast of the unconditional response mean vector $\btheta$ denoted as $f(\btheta)$,  such as the risk  difference $\theta_t - \theta_s$, risk ratio $\theta_t/\theta_s $,  and odds ratio $\frac{\theta_t/(1-\theta_t)}{\theta_s/(1-\theta_s)}$ between two treatment arms $t$ and $s$. 

Throughout the article, we consider the g-computation procedure that fits a working logistic model 
$E ( Y \mid \A, \X) = \mathrm{expit} (\bbeta_A^T \A +  \bbeta_X^T  \X )$, where $\mathrm{expit}(x) = \exp(x)/ \{1+\exp(x)\}$, and 
$\bbeta_A$ and $\bbeta_X$ are unknown parameter vectors \citep{fda:2019aa}. The  logistic model  does not need
 to be correct and is only used as an intermediate step to obtain  g-computation estimators. 
Let $\hat \bbeta_A$ and $\hat\bbeta_X$ be the maximum likelihood estimators of 
$\bbeta_A$ and $\bbeta_X$, respectively, under the working logistic model. 
Then,  $\hat \mu_t (\X_i) =    \mathrm{expit} ( \hat\bbeta_A^T \a_t + \hat \bbeta_X^T  \X_i )  $ is the predicted probability of response under treatment $t$. The g-computation estimator of  $\btheta$ is $\hat\btheta = (\hat\theta_1,\dots,  \hat\theta_k)^T$ with $\hat\theta_t = n^{-1} \sum_{i=1}^n \hat \mu_t( \X_i)$, and of a given contrast $f(\btheta)$ is $f(\hat \btheta)$. Hence, the g-computation takes a summary-then-contrast approach \citep{ICHE9R1}.
 
Next, we derive the asymptotic distribution of the g-computation estimator $\hat\btheta$ and apply the delta method to obtain  the asymptotic distribution of the g-computation estimator $f(\hat \btheta)$. As the logistic regression uses a canonical link, the first-order conditions of the maximum likelihood estimation ensure that, for $t=1, \dots, k$,  
\begin{align*}
	\sum_{i=1}^n  I(\A_i =\a_t)   \{ Y_i ^{(t)}- \hat \mu_t ( \X_i) \} = 0,
\end{align*}
where $I(\A_i=\a_t)$ is the indicator of $\A_i=\a_t$. 
Hence, the g-computation estimator is equal to
\begin{align*}
	\hat\theta_t= \frac1n \sum_{i=1}^n \hat \mu_t(\X_i)   =\frac1n \sum_{i=1}^n \left[ \frac{I(\A_i=\a_t)}{\hat \pi_{t}}  \left\{  Y_i^{(t)}  - \hat \mu_t( \X_i) \right\} + \hat \mu_t(\X_i)\right],
\end{align*}
where $\hat \pi_t = n_t/n $ and $n_t$ is the number of subjects assigned to treatment $t$.
Since $\A_i$'s are assigned completely at random, 
 $\hat \pi_t$ and  $\hat \mu_t( \bm x) $ can converge to $\pi_t$ and $\mu_t (\bm x)$ with $n^{-1/2}$ rate, respectively,  where $\bm x$ is a fixed point  and  $\mu_t(\bm x)$ is a function  not necessarily equal to  $E(Y^{(t)}\mid \X=\bm x)$ under model misspecification but 
 satisfies $E\{ Y_i^{(t)} - \mu_t (\X_i)\} = 0$ due to the above first-order conditions,  $t=1,\dots, k$. Then, by \cite{kennedy2016semiparametric} and \cite{chernozhukov2017double}, 
\begin{align*}
	\hat\theta_t = \frac1n \sum_{i=1}^n \left[ \frac{I(\A_i=\a_t)}{\pi_{t}}  \left\{  Y_i^{(t)}  -  \mu_t(\X_i) \right\} +  \mu_t( \X_i)\right] +o_p(n^{-1/2}), 
\end{align*}
where $o_p(n^{-1/2})$ denotes the remaining term multiplied by $n^{1/2}$ converges to 0  in probability.  Therefore, an application of the central limit theorem shows that, regardless of whether the working model is correct or not, 
\begin{align*}
	\sqrt{n} ( \hat \btheta - \btheta ) \xrightarrow{d} N\left( 
\bm 0, \ \bm V
	 \right), \qquad  \bm V = \begin{pmatrix} v_{11} & v_{12}& \dots& v_{1k} \\
	 	\vdots & \vdots & \ddots &\vdots \\
	 v_{1k}  & v_{2k}& \dots & v_{kk} \end{pmatrix} , 
\end{align*}
where $ \xrightarrow{d} $ denotes convergence in distribution, $\bm 0$ is the $k$-dimensional vector of zeros, and 
\begin{align*}
v_{tt} & = \pi_t^{-1}\var \{Y^{(t)} -  \mu_t(\X)\} +2 \cov \{Y^{(t)}, \mu_t(\X)\} - \var \{ \mu_t(\X)\} , \qquad t=1,\dots, k,\\
v_{ts} &= \cov \{Y^{(t)}, \mu_s(\X)\}  + \cov \{Y^{(s)}, \mu_t(\X)\} - \cov \{\mu_t(\X), \mu_s(\X)\}  , \quad 1 \leq t <s \leq k .
\end{align*} 
By the delta method, when $f(\btheta)$ is differentiable at $\btheta$ with partial derivative vector  $\nabla f(\btheta)$, we have   
\begin{align*}
	\sqrt{n} \{f(\hat\btheta) - f(\btheta)\} \xrightarrow{d} N \left(0, \ \{\nabla f(\btheta)\}^T \bm V  \{\nabla f(\btheta)\} \right). 
\end{align*} 
 Some  examples are:
 \begin{align*}
\{\nabla f(\btheta)\}^T \bm V  \{\nabla f(\btheta)\} =   \left\{ 
 		\begin{array}{ll}
 	v_{tt} - 2v_{ts}+v_{ss},  & \  f(\btheta) = \theta_t - \theta_s\\
 		\frac{v_{tt}}{\theta_t^2} - \frac{2v_{ts} }{\theta_t\theta_s} + \frac{v_{ss}}{\theta_s^2} , &  \ f(\btheta) = \log \frac{\theta_t}{\theta_s}\\
 		\frac{ v_{tt}}{\theta_t^2(1-\theta_t)^2}- \frac{2v_{ts} }{\theta_t(1-\theta_t)\theta_s(1-\theta_s) } + \frac{v_{ss}}{\theta_s^2 (1-\theta_s)^2} &  \ f(\btheta) =  \log \frac{\theta_t/(1-\theta_t)}{ \theta_s/(1-\theta_s)}.  \\
 		\end{array} \right. 
 \end{align*}
 Note that we apply normal approximation for the log transformed  risk ratio and odds ratio because the log transformation typically can improve the performance of normal approximation \citep{woolf1955estimating, haldane1956estimation}. 

For robust inference, {we propose the following variance estimator for $f(\hat\btheta)$ that is always consistent  regardless of model misspecification}:
\begin{align}
n^{-1} \{\nabla f(\hat \btheta)\}^T   \hat {\bm V} \{\nabla f(\hat\btheta)  \},
\qquad  \hat {\bm V} = \begin{pmatrix} 
	\hat v_{11} &\hat  v_{12}& \dots& \hat v_{1k} \\
	\vdots & \vdots & \ddots &\vdots \\
	\hat v_{1k}  & \hat v_{2k}& \dots &\hat  v_{kk} \end{pmatrix} ,  \label{eq: var est}
\end{align}
where 
 \begin{align*}
 \hat v_{tt}  & = \pi_t^{-1} S_{rt}^2 + 2 Q_{y tt} - S^{2}_{\mu t }, \quad t=1,\dots, k,\\
 \hat v_{ts} & = Q_{yts} + Q_{yst}  - Q_{\mu ts } , \quad 1 \leq t <s \leq k, 
 \end{align*}   $S_{rt}^2 $ is the sample variance of $Y_i - \hat \mu_t(\X_i)$ for subjects with $A_i = \a_t$, $Q_{ytt}$ is the sample covariance of $Y_i$ and $ \hat \mu_t (\X_i)$  for subjects with $A_i = \a_t$, $S_{\mu t}^2$ is the sample variance of $\hat \mu_t(\X_i)$ for all subjects,  $Q_{yts}$ is the sample covariance of $Y_i$ and $ \hat \mu_s(\X_i)$  for subjects with $A_i = \a_t$,  and $Q_{\mu ts}$ is the sample covariance of $ \hat \mu_t(\X_i)$ and $ \hat \mu_s(\X_i)$  for all subjects.
 These robust variance estimators can be directly calculated using our \textsf{R} package \textsf{RobinCar} that is publicly available at \url{https://github.com/tye27/RobinCar}.

To end this section we describe  the variance estimator in  \cite{ge2011covariate} for the g-computation estimator of 
risk difference $\hat\theta_2 - \hat\theta_1$ in a two-arm trial, and discuss why it can be inconsistent and underestimate the true variance.
 In our notation, 
 \cite{ge2011covariate} wrote the g-computation estimator  $\hat\theta_2 - \hat \theta_1  $ as $g_n (\hat \bbeta ) $, where 
 $$g_n (\hat \bbeta )  = \frac1n \sum_{i=1}^n \mathrm{expit} (\hat\bbeta_A^T \a_2 + \hat\bbeta_X^T  \X_i ) -   \frac1n \sum_{i=1}^n \mathrm{expit} (\hat\bbeta_A^T \a_1 +  \hat\bbeta_X^T  \X_i )$$
  and $\hat \bbeta = ( \hat \bbeta_A^T, \hat \bbeta_X^T)^T$. Then they applied the Taylor expansion  
  $$g_n(\hat \bbeta ) - g_n (\bbeta ) = \{  \nabla g_n ( \bbeta )\}^T (\hat \bbeta - \bbeta ) + o_p(n^{-1/2}) , $$ 
  where $\bbeta$ is the probability limit of $\hat\bbeta$, and  proposed ${n^{-1}} \{  \nabla g_n ( \hat \bbeta )\}^T  \hat {\bm V}_{\rm M} \{  \nabla g_n ( \hat \bbeta )\}  $ as a variance estimator for $\hat\theta_2 -\hat\theta_1$, where $ \hat {\bm V}_{\rm M} $ is the model-based variance estimator  for ${\sqrt{n} (\hat \bbeta - \bbeta)}$   from the standard maximum likelihood approach. This approach has  two problems. First, it uses the model-based variance estimator $  \hat {\bm V}_{\rm M} $, which may be inconsistent to the true variance of $\hat \bbeta$ under model misspecification. Second,  from 
  $$(\hat\theta_2- \hat\theta_1 ) - (\theta_2 - \theta_1) = \{ g_n(\hat \bbeta )  - g_n (\bbeta ) \} +\{  g_n (\bbeta )  - (\theta_2 - \theta_1) \} , $$
   the variance estimator proposed by \cite{ge2011covariate} 
   only accounts for the variance of the first term $  g_n(\hat \bbeta )- g_n (\bbeta ) $ but misses the variability from $ g_n (\bbeta )  - (\theta_2 - \theta_1)$ that is not 0 as the function 
   $\mathrm{expit} ( \cdot )$ is nonlinear. 
   This second problem  can lead to a confidence interval with too low coverage probability, which can be seen from the simulation results in the next section.

\section{Simulations}

We conduct simulations to evaluate the finite-sample performance of our robust variance estimator in \eqref{eq: var est}. We consider two arms or three arms, simple randomization for treatment assignments with equal allocation (i.e., $\pi_1=\pi_2=1/2$ for two arms and $\pi_1=\pi_2=\pi_3 = 1/3$ for three arms), a one-dimensional covariate  $X\sim N(0,3^2) $, and $n=200$ or $500$. 

We consider the following three  outcome data generating processes. 
 \begin{description}
	\item Case I:    $P(Y= 1\mid \A, X ) =  {\rm expit} \{-2  + 5\, 
	 I(\A= \bm a_2)+ X\}$.
	\item Case II:  $P(Y= 1\mid  \A=\a_1, X ) =  {\rm expit} (-2  + X) $ and  $P(Y= 1\mid  \A=\a_2, X ) =  {\rm expit} (3 + 1.5X - 0.01 X^2) $.
	\item Case III:   $P(Y= 1\mid \A, X ) =  {\rm expit} (-2  + 2\,
		I(\A= \bm a_2)+ 4\, 
		I(\A= \bm a_3) + X) $.
	\end{description}
 In order to determine the true values of the unconditional response means, we simulate a large dataset of  sample size $10^7$  for each case and obtain that $(\theta_1, \theta_2) = (0.2830, 0.8057) $  for Case I,  $(\theta_1, \theta_2) = (0.2830, 0.7297) $  for Case II, and $(\theta_1, \theta_2, \theta_3) = (0.2827, 0.5004, 0.7172) $  for Case III.
In each case, the g-computation estimator is based on fitting a working logistic model $P(Y=1 \mid \A , X) = {\rm expit} (\bbeta_A^T\A  + \beta_X X)$, which is correctly specified under Case I and Case III, but is misspecified under Case II.

For Case I-II, which has two arms, we focus on estimating $\theta_2- \theta_1$ and also  include the variance estimator in  \cite{ge2011covariate}. For Case III, which has three arms, we evaluate our robust variance estimators for three common unconditional treatment effects for binary outcomes.  The results for Case I-II are in Table \ref{tb} and for Case III are in Table \ref{tb2}, which include (i) the  true parameter value, (ii) Monte Carlo mean  and standard deviation (SD) of g-computation point estimators, (iii) average of standard error (SE); and (iv) coverage probability (CP) of 95\% confidence intervals. We use sample size $n =$ 200 or 500, and 10,000 simulation runs.

From Tables \ref{tb}-\ref{tb2}, we see that the g-computation estimators have negligible biases compared to the standard deviations. Our robust standard error, which is the squared root of variance estimator in \eqref{eq: var est}, is always very close to the actual standard deviation, and the related confidence interval has nominal coverage across all settings.  In contrast, the standard error in \cite{ge2011covariate} underestimates the actual standard deviation under Case I when there is no model misspecification, as well as under Case II when there is   model misspecification, and the related confidence intervals have too low coverage probabilities in both cases.

\vspace{5mm}

\begin{table}[h]
	\centering
	\caption{Simulation mean and standard deviation (SD) of $\hat\theta_2-\hat\theta_1$, average standard error (SE), and coverage probability (CP) of 95\% asymptotic confidence interval for $\theta_2 - \theta_1$ under Case I-II and simple randomization. \label{tb}}
	\begin{tabular}{ccccccccccc}  \hline 
 &		& & \multicolumn{2}{c}{$\hat\theta_2-\hat\theta_1$} & &
		\multicolumn{2}{c}{Robust SE in \eqref{eq: var est} } 
&		& \multicolumn{2}{c}{SE in \cite{ge2011covariate} }\\ \cline{4-5} \cline{7-8} \cline{10-11} 
Case		& $\theta_2-\theta_1$   & $n$	         & Mean      & SD           &           & SE     & CP  (\%) && SE & CP  (\%) \\ \hline 
I &   0.5227  & 200 & 0.5228 & 0.0464 & & 0.0464 & 94.44 & & 0.0415   &91.27        \\
& &500    & 0.5227  & 0.0295 & & 0.0294 & 94.70 & & 0.0264 & 91.94 \\
II       &   0.4467 & 200     &         0.4469    &0.0457 &   & 0.0458 & 94.56 && 0.0404 & 91.08 \\
& & 500 &  {0.4463}   & 0.0289 && 0.0290 & 94.90  &               & 0.0257 & 91.77 \\\hline 
	\end{tabular}
\end{table}

\begin{table}[h]
	\centering
	\caption{Simulation mean and standard deviation (SD) of g-computation estimators, average standard error (SE), and coverage probability (CP) of 95\% asymptotic confidence interval based on robust SE (\ref{eq: var est}) under Case III and simple randomization. \label{tb2}}
	\begin{tabular}{ccccccccccc} \hline
		&             & \multicolumn{4}{c}{$n=200$}      &  & \multicolumn{4}{c}{$n=500$}        \\  \cline{3-6}  \cline{8-11}  
Parameter		& Truth       & Mean   & SD     & SE     & CP&    & Mean   & SD     & SE     & CP    \\ \hline 
	$\theta_2-\theta_1$ & 0.2177 & 0.2176 & 0.0578 & 0.0573 & 94.34 &  & 0.2170 & 0.0366 & 0.0363 & 94.82 \\
		$\log (\theta_2/\theta_1)$ & 0.5711 & 0.5798 & 0.1701 & 0.1664 & 94.50 &  & 0.5726 & 0.1053 & 0.1042 & 94.59 \\
	$ \log \frac{\theta_2/(1-\theta_2)}{ \theta_1/(1-\theta_1)}$ & 0.9328 & 0.9440 & 0.2620 & 0.2586 & 94.63 &  & 0.9341 & 0.1637 & 0.1624 & 94.79 \\
&        &        &        &        &       &  &        &        &        &       \\
		$\theta_3-\theta_1$   & 0.4346 & 0.4348 & 0.0581 & 0.0568 & 94.15 &  & 0.4347 & 0.0360 & 0.0360 & 94.84 \\
	$\log (\theta_3/\theta_1)$ & 0.9311 & 0.9432 & 0.1653 & 0.1611 & 94.43 &  & 0.9353 & 0.1018 & 0.1009 & 94.92 \\
$ \log \frac{\theta_3/(1-\theta_3)}{ \theta_1/(1-\theta_1)}$ & 1.8621 & 1.8852 & 0.2920 & 0.2851 & 94.57 &  & 1.8712 & 0.1791 & 0.1788 & 95.01\\\hline 
	\end{tabular} 
\end{table}

\vspace{5mm}

\section{Summary and Discussion}

In this article, we provide an explicit robust variance estimator formula for g-computation estimators, which can be used for different unconditional treatment effects of interest and clinical trials with two or more arms. Our simulations demonstrate that the variance estimator can be reliably used in practice. 

In this article, for the purpose of being specific, we focus on the logistic model that regresses the outcome on the treatment indicators and covariates, which is arguably the most widely used model for binary outcomes. However, our robust variance estimation formula in \eqref{eq: var est} is not limited to this model and 
can be used with different specifications of the working model  (e.g., fitting a separate logistic model for each treatment arm) or with other generalized linear models using a canonical link for non-binary outcomes (e.g., Poisson regression for count outcomes). Additionally,   although our article considers simple
 randomization,	 our robust variance formula in \eqref{eq: var est} can also be used for  
	a 
	complete randomization scheme where the sample size in every group $t$ is fixed to be $n\pi_t$,
	 because this randomization scheme leads to the same asymptotic distribution as the simple randomization \citep{ye2021better}. Simulation results under this randomization scheme  are similar to those under simple randomization; see	 Tables 3-4 in the Appendix.

We implement an \textsf{R} package called   \textsf{RobinCar}
to conveniently compute the g-computation estimator and our robust variance estimators, which is publicly available at \url{https://github.com/tye27/RobinCar}.

\begin{center}
	{\sffamily\bfseries\LARGE
		Appendix
	}
\end{center}

In Tables \ref{tb3}-\ref{tb4}, we include simulation results under a complete randomization scheme where the sample size in every group $t$ is fixed to be $n\pi_t$.

\begin{table}[h]
	\centering
	\caption{Simulation mean and standard deviation (SD) of $\hat\theta_2-\hat\theta_1$, average standard error (SE), and coverage probability (CP) of 95\% asymptotic confidence interval for $\theta_2 - \theta_1$ under Case I-II and complete randomization that fixes $n_t= n\pi_t$. \label{tb3}}
	\begin{tabular}{ccccccccccc}  \hline 
		&		& & \multicolumn{2}{c}{$\hat\theta_2-\hat\theta_1$} & &
		\multicolumn{2}{c}{Robust SE in \eqref{eq: var est} } 
		&		& \multicolumn{2}{c}{SE in \cite{ge2011covariate} }\\ \cline{4-5} \cline{7-8} \cline{10-11} 
		Case		& $\theta_2-\theta_1$   & $n$	         & Mean      & SD           &           & SE     & CP  (\%) && SE & CP  (\%) \\ \hline 
		I &   0.5227  & 200  & 0.5231 & 0.0464 && 0.0462 & 94.62 && 0.0414 & 91.54 \\
		& &500   &  0.5230 & 0.0298 && 0.0294 & 94.67 && 0.0264 & 91.70 \\
		II       &   0.4467 & 200     &        0.4469 & 0.0457 && 0.0456 & 94.68 && 0.0403 & 91.29 \\
		& & 500 & 0.4471 & 0.0290 && 0.0290 & 94.74 && 0.0257 & 91.45  \\\hline 
	\end{tabular}
\end{table}

\begin{table}[h]
	\centering
	\caption{Simulation mean and standard deviation (SD) of g-computation estimators, average standard error (SE), and coverage probability (CP) of 95\% asymptotic confidence interval based on robust SE (\ref{eq: var est}) under Case III and  complete randomization that fixes $n_t= n\pi_t$. \label{tb4}}
	\begin{tabular}{ccccccccccc} \hline
		&             & \multicolumn{4}{c}{$n=200$}      &  & \multicolumn{4}{c}{$n=500$}        \\  \cline{3-6}  \cline{8-11}  
		Parameter		& Truth       & Mean   & SD     & SE     & CP&    & Mean   & SD     & SE     & CP    \\ \hline 
		$\theta_2-\theta_1$ & 0.2177 & 0.2177 & 0.0580 & 0.0570 & 94.25 && 0.2182 & 0.0364 & 0.0362 & 94.61  \\
		$\log (\theta_2/\theta_1)$ & 0.5711 & 0.5790 & 0.1687 & 0.1652 & 94.60 && 0.5755 & 0.1046 & 0.1040 & 95.02  \\
		$ \log \frac{\theta_2/(1-\theta_2)}{ \theta_1/(1-\theta_1)}$ & 0.9328 & 0.9437 & 0.2616 & 0.2567 & 94.78 && 0.9391 & 0.1627 & 0.1621 & 94.73 \\
		&        &        &        &        &       &  &        &        &        &       \\
		$\theta_3-\theta_1$   & 0.4346 &0.4349 & 0.0579 & 0.0567 & 93.96 && 0.4354 & 0.0364 & 0.0360 & 94.23 \\
		$\log (\theta_3/\theta_1)$ & 0.9311 & 0.9424 & 0.1640 & 0.1602 & 94.02 && 0.9371 & 0.1017 & 0.1008 & 94.99 \\
		$ \log \frac{\theta_3/(1-\theta_3)}{ \theta_1/(1-\theta_1)}$ & 1.8621 & 1.8853 & 0.2910 & 0.2844 & 94.31 & & 1.8751 & 0.1811 & 0.1788 & 94.48\\\hline 
	\end{tabular} 
\end{table}

\bibliographystyle{apalike}
\bibliography{reference}

\end{document}